# Correct Asymptotic Wavefunctions for Calculating Photoelectron Angular Distributions of $O_2^-$ and $NO^-$

Wenru Jie, Rui Zhang, Jiayi Chen, Qihan Liu, and Chuangang Ning[*]

*Department of Physics, State Key Laboratory of Low Dimensional Quantum Physics, Frontier Science Center for Quantum Information, Tsinghua University, Beijing 100084, China*

[*] Author to whom correspondence should be addressed: ningcg@tsinghua.edu.cn

## ABSTRACT

The *ab initio* calculation of photoelectron angular distributions (PADs) for negative ions remains a significant theoretical challenge. In this work, we report a joint experimental and theoretical investigation of PADs for a series of molecular anions with varying polarities, including the nonpolar $O_2^-$, the weakly polar $NO^-$, and the strongly polar $AsO^-$ and $SbO^-$. To accurately describe the long-range electronic wavefunctions—where photodetachment contributes most strongly—we modified the standard Gaussian-type orbitals (GTOs) by augmenting them with a correct exponential Slater-tail basis set ($\sim e^{-\xi r}$). This simple yet effective approach significantly improves the agreement between the experimental and theoretical PADs for $O_2^-$ and $NO^-$. However, notable discrepancies persist for $NO^-$ for transitions to the $v = 0$ and $v = 1$ vibrational levels of neutral NO even after this correction. Given that our methodology successfully reproduced PADs for strongly polar anions (e.g., $AsO^-$ and $SbO^-$), these residual discrepancies are unlikely to stem from "exit-channel scattering" induced by long-range dipole fields. Instead, we tentatively attribute the failure for $NO^-$ to the breakdown of the Born-Oppenheimer approximation or the frozen orbital approximation, arising from the extremely weak binding of the excess electron.

## I. INTRODUCTION

Photoelectron spectroscopy (PES) based on the velocity-map imaging (VMI) method, particularly its cryogenic slow-electron variant (cryo-SEVI), provides high-resolution energy spectra and photoelectron angular distributions (PADs) simultaneously. Compared to the fruitful information extracted from the energy spectra [1–5], our understanding of PADs remains limited [6–8]. The photoelectron angular distributions (PADs), generally characterized by the anisotropy parameter $\beta$, are sensitive to the electronic wavefunctions and the kinetic energy of photoelectrons. While the central-potential model established by Cooper and Zare [9,10] and extended analytical models [7,11–16] succeed for certain simple systems, their reliance on empirical parameter fitting limits their predictive power.

Several theoretical methods have been developed for calculating PADs from first principles [6,8,16–18], which have significantly improved our understanding of PADs. Despite this progress, notable discrepancies between the calculated and experimental PADs still exist. A prominent example is the inability of conventional *ab initio* calculations to reproduce the vibrational dependence observed in the PADs of homonuclear nonpolar molecules like $O_2^-$ [14,19]. For polar molecules, the exit-channel scattering model incorporates long-range electron-molecule interactions to address these predictive limitations [17,18]. Nevertheless, the calculated $\beta$ parameters still show significant deviations from the experimental measurements.

To identify the origin of these discrepancies, it is instructive to compare systems spanning a broad range of molecular polarities and electron-binding energies. In this work, we investigate three distinct types of molecular anion systems: the nonpolar $O_2^-$, the weakly polar $NO^-$, and the strongly polar $AsO^-$ and $SbO^-$. Our previous work [6] has shown that the PAD calculations based on Slater-type orbitals (STOs) perform better than those based on Gaussian-type orbitals (GTOs) because STOs have the correct asymptotic form $\sim e^{-\xi r}$ at long range [20,21]. The parameter $\xi$ is related to the electron binding energy eBE of the detached state: $\xi = \sqrt{2m_e \cdot \mathrm{eBE}}/\hbar$. Because the electron binding energies involved in negative ions are typically small (~1 eV), the diffuse tail of the bound-state wavefunction extends far from the molecular core. In this long-range region, GTOs decay as $\sim e^{-\alpha r^2}$, which is significantly faster than the exponential form $\sim e^{-\xi r}$. The importance of this asymptotic region is further enhanced by the electric-dipole transition operator. Because the dipole operator contains a radial weighting factor proportional to $r$, contributions from the long-range region are enhanced in the photodetachment amplitude, making PADs particularly sensitive to the asymptotic behavior of the initial-state wavefunction. To address these limitations, we develop a Slater-tail augmented basis set that improves the asymptotic behavior of the initial-state wavefunction. We apply this framework to the PADs of $O_2^-$, $NO^-$, $AsO^-$, and $SbO^-$, providing a systematic comparison across molecular anions with markedly different polarities and binding energies. By comparing experimental measurements with theoretical simulations, we demonstrate that an accurate description of the long-range bound-state wavefunction can substantially improve the agreement between theoretical PADs and experimental results.

## II. METHODS

### A. Experimental methods

The experiments were performed using our home-built cryo-SEVI apparatus [22–24]. Anion generation techniques varied depending on the target system: $NO^-$ and $O_2^-$ were produced by expanding $N_2O$ gas (~3 bar) through a pulsed valve equipped with an electron-gun ion source, whereas $AsO^-$ and $SbO^-$ were generated via laser ablation of As and Sb targets in the $N_2O$ carrier gas. The generated anions were guided by a radio frequency (RF) hexapole into an RF octupole ion trap maintained at 7.5 K. After a 45 ms trapping period, the anions were cooled to their vibrational ground states and a few rotational excited states through collisions with a buffer gas mixture (20% $H_2$ + 80% He).

The anions of interest were mass-selected using a time-of-flight mass spectrometer and subsequently photodetached by tunable laser irradiation. The outgoing photoelectrons were mapped onto a microchannel plate/phosphor screen assembly and recorded by a CCD camera. A typical 2D photoelectron image was accumulated over approximately 50,000 laser shots. The 3D velocity distributions can be reconstructed using the MEVELER algorithm from the projected 2D image [25]. As shown in Fig.1, a series of experimental energy spectra and the photoelectron angular distribution images are acquired by varying the laser wavelength. The extracted $\beta$ values were then plotted versus the photoelectron kinetic energy.

### B. Theoretical and Computational Methods

To calculate PADs and $\beta$, we adopted the theoretical framework developed by Liu and Ning [6]. Within the electric-dipole approximation, photodetachment is described by the transition matrix element $M_k^{fi} = \langle \Psi_f^{N-1}\Psi_k \mid \vec{E} \cdot \vec{r} \mid \Psi_i^N \rangle$, where $\Psi_i^N$ is the initial anion state, $\Psi_f^{N-1}$ is the final neutral core state, $\Psi_k$ is the free photoelectron wavefunction with wavevector *k*, and *r* is the dipole operator. In practice, PADs were evaluated directly from the highest occupied molecular orbital (HOMO) of the anion within the frozen orbital approximation [26]. Because the transition dipole operator contains an explicit radial factor *r*, contributions from the asymptotic region are strongly amplified in the matrix element. As a result, even small inaccuracies in the long-range wavefunction can significantly affect the predicted PADs.

To accurately represent the long-range behavior of the electron wavefunction in photodetachment, we modified Gaussian-type orbital (GTO) basis sets. The construction procedure and criteria for the basis sets are as follows:

1. Even-tempered extension and reorganization of diffuse functions

The standard aug-cc-pV5Z basis set consists of contracted blocks describing the core region and several uncontracted diffuse functions. First, the most diffuse Gaussian exponent from the contracted primitive basis set was extracted as the starting point. Subsequently, an even-tempered sequence ($\alpha_{i+2} = \alpha_{i+1}^2/\alpha_i$) was employed to extend the set towards smaller exponents, generating a series of new diffuse Gaussian functions. Finally, the original uncontracted diffuse functions were combined with the newly generated ones to construct an auxiliary contracted basis set block.

2. Asymptotic behavior fitting

The asymptotic behavior of the long-range wavefunction was incorporated as a constraint in the basis-set construction. An auxiliary contracted basis function, $\psi_{ST}(r)$, was expressed as a linear combination of diffuse Gaussian primitives,

$$\psi_{ST}(r) = \sum_i c_i e^{-\alpha_i r^2} \tag{1}$$

where $\alpha_i$ are the Gaussian exponents and $c_i$ are the contraction coefficients. The coefficients $c_i$ were determined through a least-squares fitting procedure over the radial range (1.2 Å < r < 20 Å), such that the contracted Gaussian combination reproduces the target Slater-type asymptotic decay ($\sim e^{-\xi r}$). In this way, the modified basis set preserves the computational advantages of GTOs while accurately reproducing the physically correct exponential behavior in the asymptotic region.

3. State-dependent decay parameter

The asymptotic decay parameter $\xi$ was calibrated according to the binding energy and spatial extent of different vibrational states for each system (e.g., $\xi = 0.27$ a.u.$^{-1}$ for the $v = 0$ vibrational state of $O_2^-$). After determining the contraction coefficients $c_i$ through fitting, 1–2 of the most diffuse primitive Gaussian functions from the original

aug-cc-pV5Z basis were retained as independent uncontracted functions (coefficient = 1.0). This procedure prevents the fitted Slater-tail block from becoming overly rigid and preserves sufficient variational flexibility during the self-consistent-field (SCF) optimization. It also improves numerical stability in subsequent electronic-structure calculations.

**Fig.2** compares the asymptotic forms of the HOMO of $O_2^-$ in the long range generated using different basis sets. The curves using the conventional aug-cc-pVTZ and aug-cc-pV5Z basis sets decay too rapidly in the long range, while the present Slater-tail basis set correctly reproduces the expected linear behavior on a logarithmic scale. The 3D orbital in Fig.2 is generated using Multiwfn [27] and rendered with VMD [28].

# III. RESULTS AND DISCUSSION

## A. Vibrationally Resolved PADs of $O_2^-$

**FIG. 4** shows the experimentally measured anisotropy parameter $\beta$ as a function of electron kinetic energy for different vibrational levels of the corresponding neutral product ($v$ = 0–4) in comparison with the theoretical calculations. As shown in **FIG. 4**(a), standard Gaussian basis sets (e.g., aug-cc-pVTZ and aug-cc-pV5Z) exhibit deviations in predicting the kinetic energy dependence of $\beta$. This confirms that standard GTOs, due to their rapid asymptotic decay, cannot correctly describe the long-range wavefunction tail which is crucial for determining the photodetachment dynamics.

In contrast, using the Slater-tail basis set, our theoretical calculations agree with the experiments for all vibrational channels (**FIG. 4**(b)–(f)). To incorporate these vibrational dynamics into our theoretical model, we adjusted two vibrational-state-specific parameters for each transition channel: the asymptotic decay parameter ($\xi$) and the effective internuclear distance ($d$). First, because vibrational states modify the effective binding energy [14] , we directly used these binding energies to determine $\xi$ for the asymptotic tail of each wavefunction. **Table 1** lists the physical properties and the $\xi$ values employed for all systems. Following this approach, the calculated $\xi$ increases systematically from $\xi = 0.27$ a.u.$^{-1}$ at $v = 0$ to $\xi = 0.32$ a.u.$^{-1}$ at $v = 4$. Second, to account for the different vibrational states, the effective internuclear distance $d$ was chosen to correspond to the maximum spatial overlap between the ground vibrational wavefunction of the initial anion and the final neutral molecule for the specific vibrational level.

## A. Vibrationally Resolved PADs of Weakly Bound $NO^-$

For the weakly polar anion $NO^-$, whose neutral core possesses a dipole moment of ~0.158 Debye, the experimental observations present a different dynamical evolution. As shown in **FIG. 1**, the $\beta$ curves for the $v = 0$ and $v = 1$ states of $NO^-$ are flat, contrasting with the steep decline observed in $O_2^-$. We applied the Slater-tail model to this system. For the higher vibrational states (e.g., $v = 2$, 3, and 4), the theoretical calculations agree well with the experimental evolutionary trends. However, for the $v = 0$ and $v = 1$ states, the calculations show significant deviations from the experimental observations, particularly at higher kinetic energies.

These deviations prompted us to examine whether the discrepancy originates from the polarity of the NO molecule. In a polar molecular anion, the detached free electron is scattered by the dipole field as it leaves the core region, which is conventionally called the "exit-channel scattering" effect [17,18]. This motivated further investigation into the impact of dipole-field effects on photodetachment dynamics.

## B. PADs of $AsO^-$ and $SbO^-$

To test the polarity hypothesis, we investigated the more polar metal oxide systems $AsO^-$ and $SbO^-$ in the same group. These two molecules possess permanent dipole moments of 2.7 Debye and 3.6 Debye, respectively, which far exceed that of NO. As shown in **FIG. 6**, despite their larger dipole moments, their PADs exhibit regular features, and our calculated PADs agree well with the experimental results. Because the total binding energies of $AsO^-$ and $SbO^-$ are substantially larger than their respective vibrational or spin-orbit energy splittings, the relative variation in $\xi$ between these adjacent sub-states is negligible. Consequently, the Slater-tail model reproduces the experimental trends for all observed states using a single $\xi$ parameter (e.g., $\xi \approx 0.24$ a.u.$^{-1}$ for $AsO^-$, $\xi \approx 0.22$ a.u.$^{-1}$ for $SbO^-$).

The success of our method in reproducing the PADs of these strongly polar systems—which should theoretically be far more sensitive to dipole scattering—suggests that exit-channel scattering is unlikely to be the dominant factor for the observed discrepancy in the photodetachment of $NO^-$. One possible origin of the discrepancies observed in $NO^-$ is its extremely low electron affinity. The electron affinity (EA) of NO (235.8 $cm^{-1}$) is one order of magnitude smaller than its vibrational energy (1881.04 $cm^{-1}$) [29]. In this regime, electronic and nuclear motions may no longer be fully separable within the Born-Oppenheimer approximation. Another possible reason is the failure of the frozen orbital approximation, which assumes that after the photodetachment of an electron from the HOMO, the other orbitals remain unperturbed.

## IV. CONCLUSIONS

In summary, this work presents a systematic investigation of vibrational-state-dependent photoelectron angular distributions (PADs) for $O_2^-$ and $NO^-$ across a range of photon energies. By implementing a Slater-tail augmented basis set, we effectively mitigate the unphysically rapid decay inherent in conventional Gaussian-type orbitals. This modification yields substantially improved agreement between theoretical predictions and experimental measurements, underscoring the critical importance of capturing the correct asymptotic behavior of anion wavefunctions in PAD calculation.

Nevertheless, significant discrepancies persist for the lowest vibrational channels for the weakly bound $NO^-$ system. The deviations are unlikely to be attributable to exit-channel scattering via long-range dipole fields, as our methodology successfully reproduced PADs for the strongly polar anions $AsO^-$ and $SbO^-$ within the same group. We tentatively attribute the observed failure for $NO^-$ to the breakdown of either the Born-Oppenheimer approximation or the frozen-orbital approximation, stemming from the extremely weak binding of the excess electron.

## ACKNOWLEDGEMENTS

This work was supported by the National Natural Science Foundation of China (NSFC) (Grant Nos. 12374244 and 12341401).

## DATA AVAILABILITY

The data that support the findings of this article are openly available.

**Table 1** Summary of the relevant physical properties (electron affinity EA, vibrational spacing $\omega_e$, and permanent dipole moment) and the corresponding asymptotic decay parameters $\xi$ employed in the Slater-tail basis set calculations for the investigated molecular anions.

| System | $O_2$ | AsO | SbO | NO |
|---|---|---|---|---|
| EA ($cm^{-1}$) | 3611.0(80) [a] | 10318(7) [b] | 12426(7) [b] | 235.8(40) [a] |
| $\omega_e$ ($cm^{-1}$) | 1580.19 [c] | 966.49 [c] | 858 [d] | 1904.20 [c] |
| Dipole Moment (Debye) | 0 | 2.70 [e] | 3.6 [d] | 0.16 [f] |
| Parameter $\xi$ (a.u.$^{-1}$) | 0.27 for $v = 0$<br>0.287 for $v = 1$<br>0.30 for $v = 2$<br>0.31 for $v = 3$<br>0.32 for $v = 4$ | 0.24 | 0.22 | 0.195 for $v = 0$<br>0.21 for $v = 1$<br>0.24 for $v = 2$<br>0.26 for $v = 3$<br>0.28 for $v = 4$ |

Notes: a) From Ref. [5]

b) From experiment in this work.

c) From Ref. [30]

d) Theoretically calculated in this work at the UB3LYP/GenECP level. The Sb atom was described using the aug-cc-pV5Z-PP small-core relativistic pseudopotential [31–33] which replaces 28 inner-shell core electrons ([Ar]$3d^{10}$), leaving 23 valence electrons to be treated explicitly. The valence space of both Sb and O atoms was described by our custom Slater-tail modified basis sets built upon the aug-cc-pV5Z framework.

e) From Ref. [34]

f) From Ref. [35]

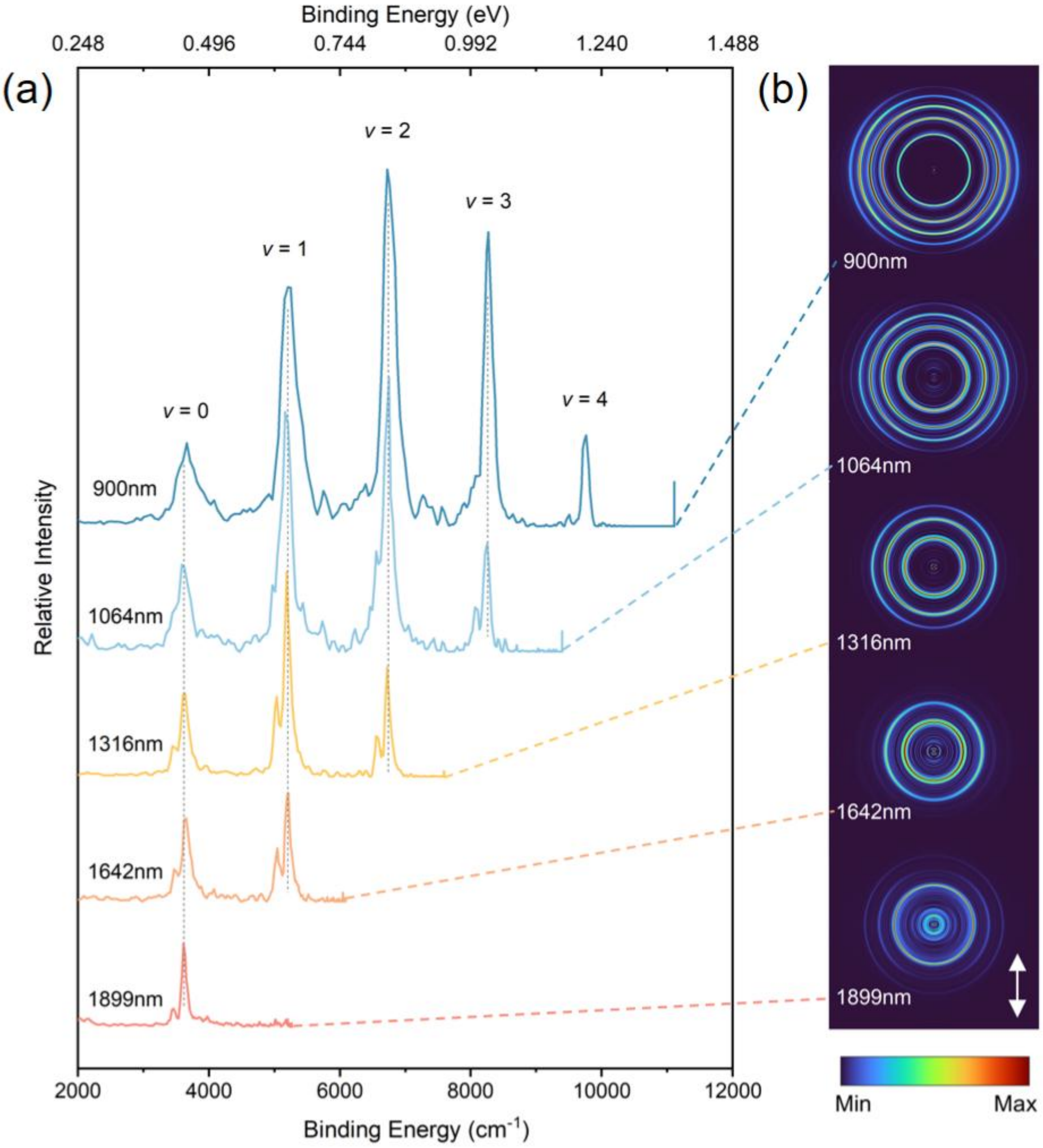

**FIG. 2.** Schematic of the experimental method using $O_2$ as an example. (a) Photoelectron energy spectra of $O_2^-$ with neutral $O_2$ at different vibrational states ($v$ = 0–4) recorded at various laser wavelengths. (b) Corresponding angular distribution images acquired at each wavelength.

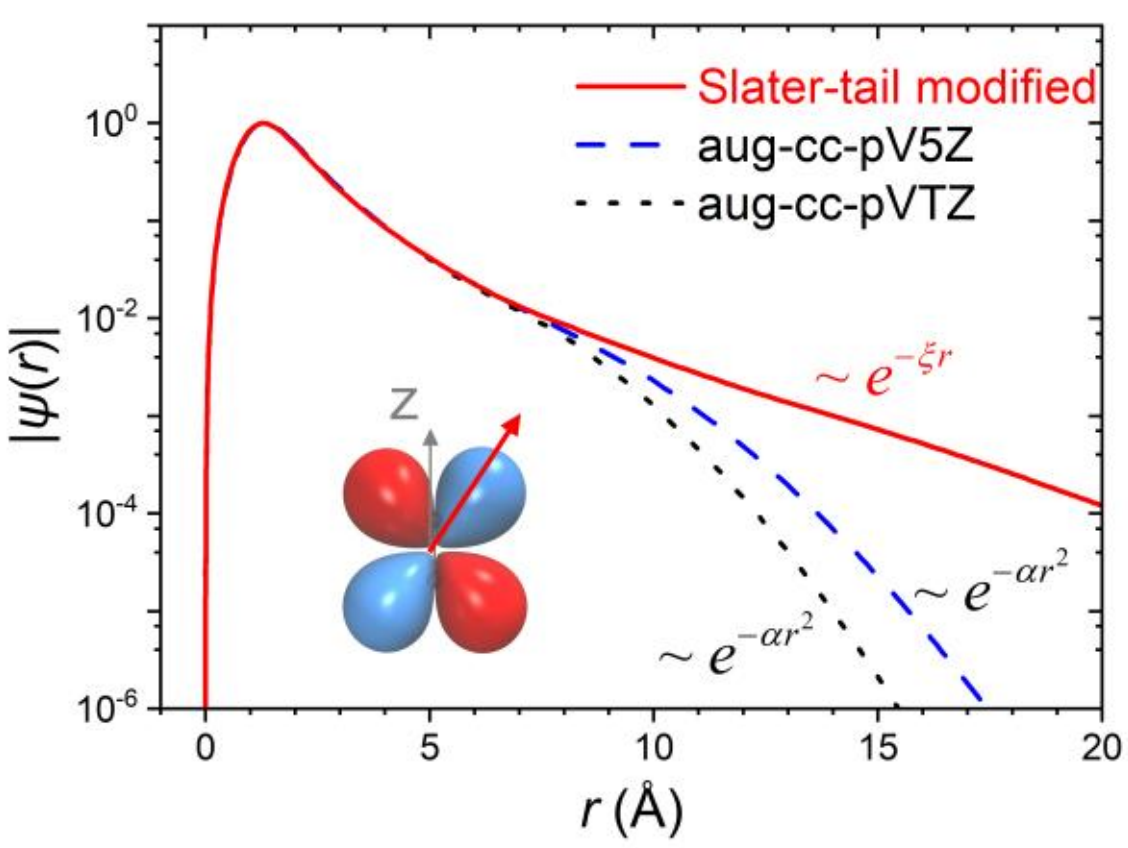

**FIG. 3.** Comparison of radial wavefunctions of the highest occupied molecular orbital (HOMO) of $O_2^-$ calculated using different theoretical methods and basis sets. As illustrated in the 3D orbital inset, the 1D wavefunctions are extracted along the specific radial direction indicated by the red arrow. These wavefunctions are compared on a logarithmic scale to emphasize their asymptotic tail behaviors. In this representation, a Slater-type tail ($\sim e^{-\xi r}$) appears as a straight line, whereas a Gaussian-type tail ($\sim e^{-ar^2}$) appears as a parabola.

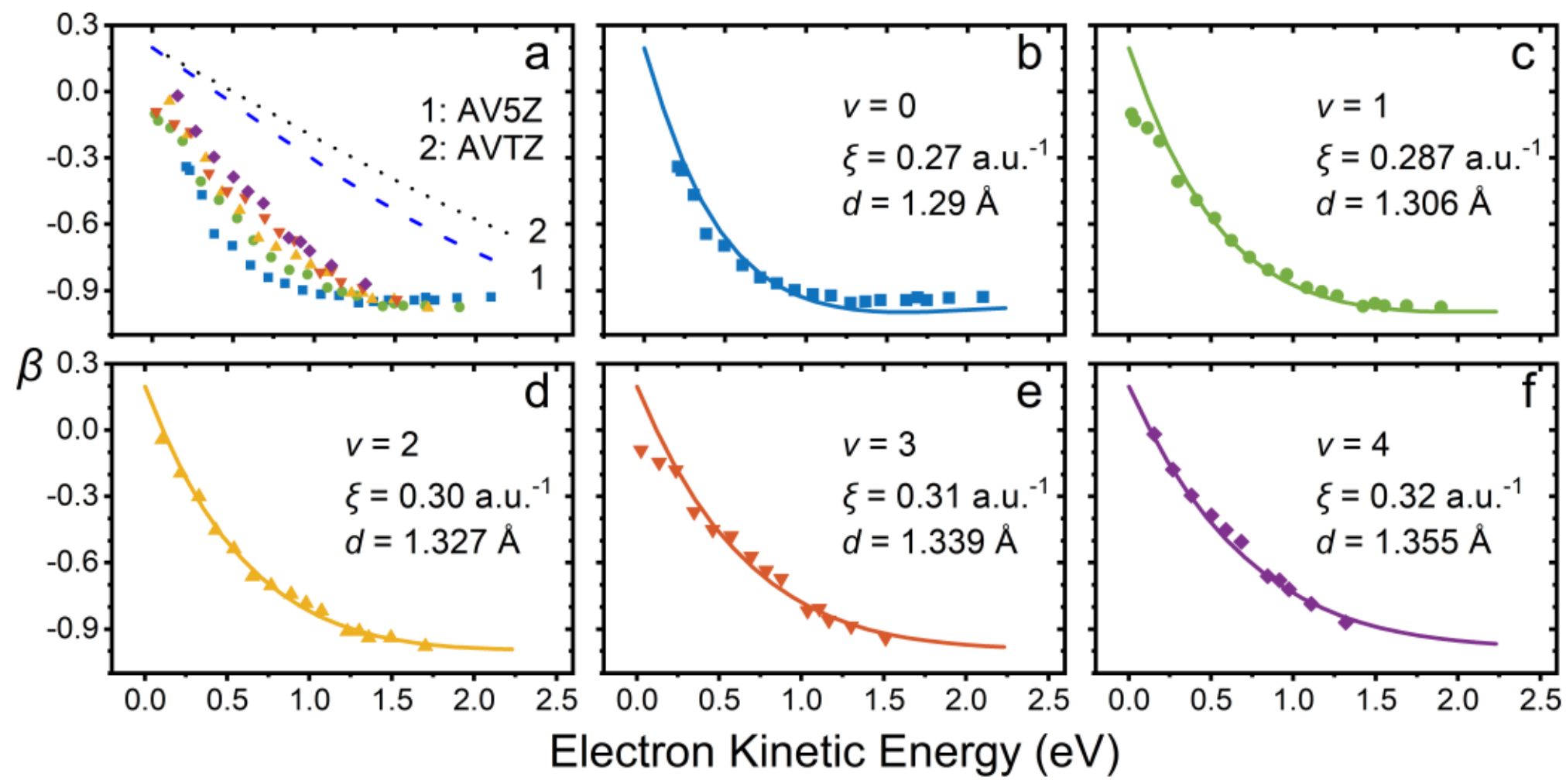

**FIG. 4.** Anisotropy parameter $\beta$ for photodetachment from $O_2^-$ as a function of electron kinetic energy for each vibrational level of the neutral $O_2$ product. (a) Summary of all experimental $\beta$ values for $v$ = 0–4 together with theoretical results obtained using the aug-cc-pVTZ (dotted line) and aug-cc-pV5Z (dashed line) basis sets. (b)–(f) Detailed comparison between experimental data and theoretical calculations using the Slater-tail basis set. The theoretical curves are obtained with the asymptotic parameter $\xi$ and effective internuclear distance $d$ indicated in each panel.

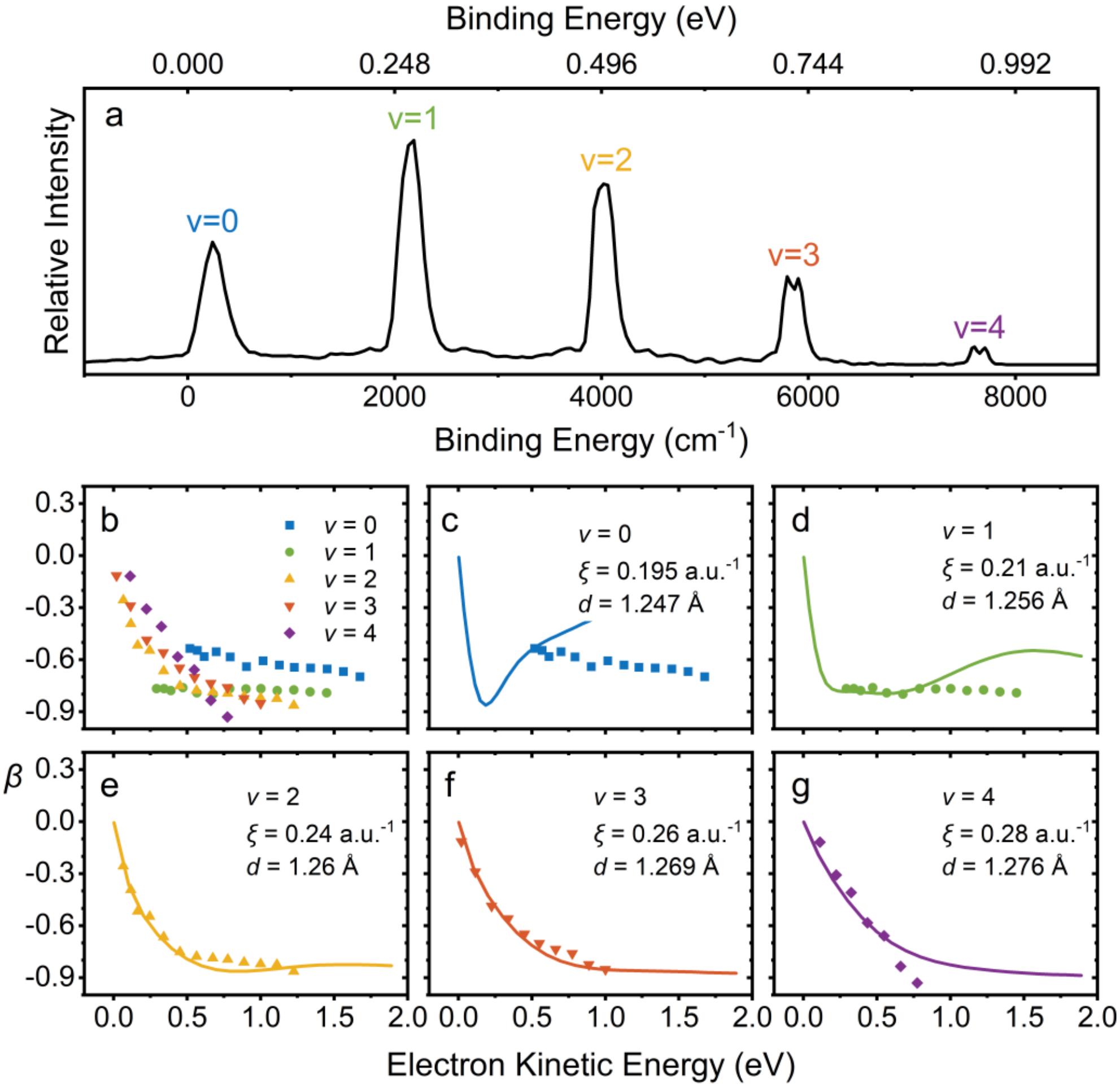

**FIG. 5.** Anisotropy parameter $\beta$ for photodetachment from $NO^-$ as a function of electron kinetic energy for each vibrational level of the neutral NO product. (a) Photoelectron energy spectrum of $NO^-$ recorded with a laser wavelength of 1064 nm. (b) Summary of all experimental $\beta$ values for $v$ = 0–4. (c)–(g) Detailed comparison between experimental data and theoretical calculations for transitions to $v$ = 0–4, respectively.

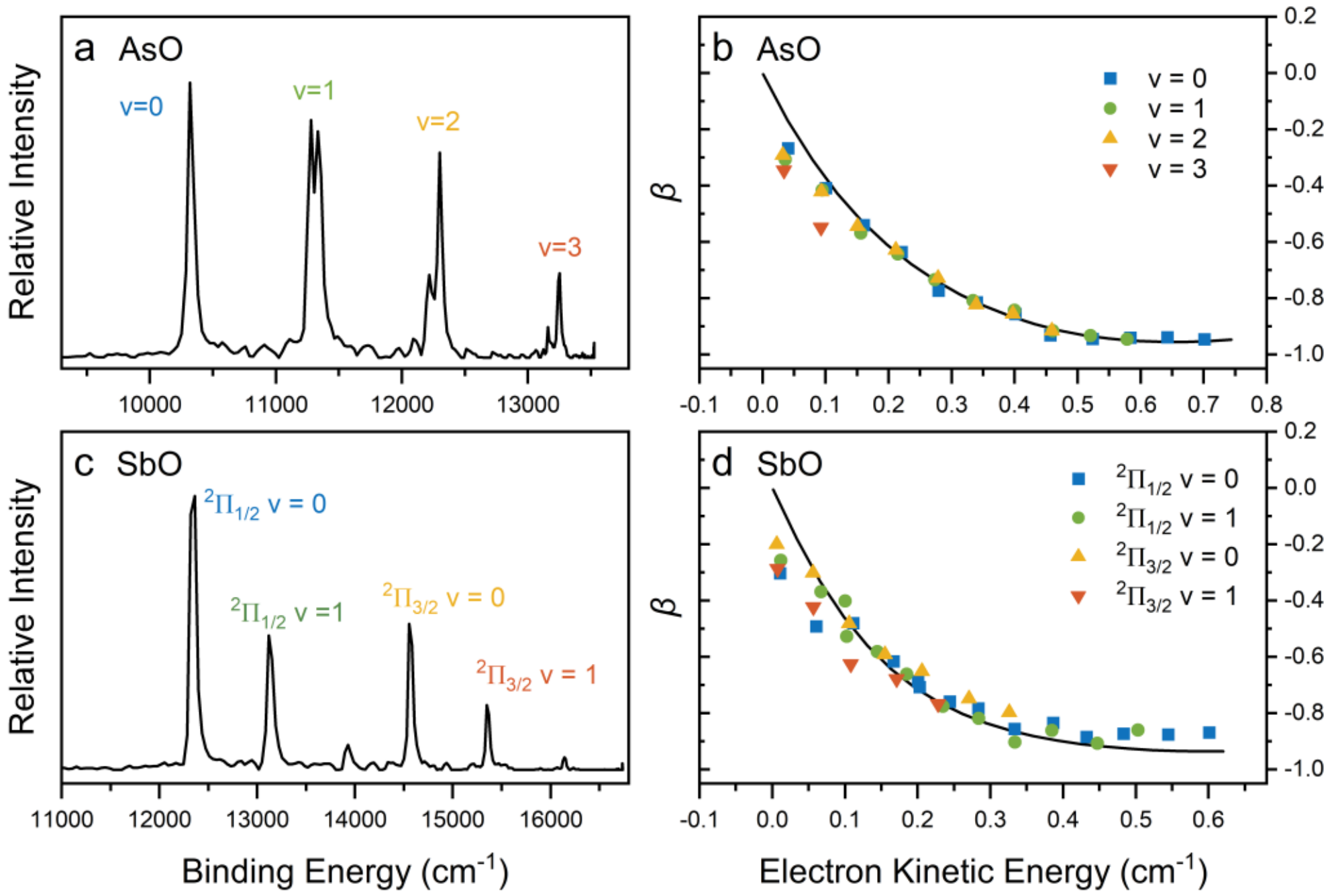

**FIG. 6.** Anisotropy parameter $\beta$ for photodetachment from $AsO^-$ and $SbO^-$ as a function of electron kinetic energy. (a) Photoelectron energy spectrum of $AsO^-$ recorded with a laser wavelength of 739.37 nm. (b) Corresponding $\beta$ values for AsO as a function of electron kinetic energy. (c) Photoelectron energy spectrum of $SbO^-$ recorded with a laser wavelength of 601 nm. (d) Corresponding $\beta$ values for SbO.

## Reference

[1] A. Osterwalder, M. J. Nee, J. Zhou, and D. M. Neumark, High resolution photodetachment spectroscopy of negative ions via slow photoelectron imaging, J. Chem. Phys. **121**, 6317 (2004).

[2] L.-S. Wang, Perspective: Electrospray photoelectron spectroscopy: From multiply-charged anions to ultracold anions, J. Chem. Phys. **143**, 040901 (2015).

[3] M. L. Weichman and D. M. Neumark, Slow Photoelectron Velocity-Map Imaging of Cryogenically Cooled Anions, Annu. Rev. Phys. Chem. **69**, 101 (2018).

[4] R. L. Tang, R. Si, Z. J. Fei, X. X. Fu, Y. Z. Lu, T. Brage, H. Liu, C. Chen, and C. Ning, Candidate for Laser Cooling of a Negative Ion: High-Resolution Photoelectron Imaging of $Th^-$, Phys. Rev. Lett. **123**, 203002 (2019).

[5] W. R. Jie, J. Y. Chen, R. Zhang, S. T. Yan, and C. G. Ning, Rotational envelope simulations in photodetachment spectroscopy: Precise measurement of electron affinity of NO and $O_2$, J. Chem. Phys. **163**, 044309 (2025).

[6] Y. Liu and C. Ning, Calculation of photodetachment cross sections and photoelectron angular distributions of negative ions using density functional theory, J. Chem. Phys. **143**, 144310 (2015).

[7] A. Sanov, Laboratory-Frame Photoelectron Angular Distributions in Anion Photodetachment: Insight into Electronic Structure and Intermolecular Interactions, Annu. Rev. Phys. Chem. **65**, 341 (2014).

[8] C. M. Oana and A. I. Krylov, Cross sections and photoelectron angular distributions in photodetachment from negative ions using equation-of-motion coupled-cluster Dyson orbitals, J. Chem. Phys. **131**, 124114 (2009).

[9] J. Cooper and R. N. Zare, Angular Distribution of Photoelectrons, J. Chem. Phys. **48**, 942 (1968).

[10] J. Cooper and R. N. Zare, Erratum: Angular Distribution of Photoelectrons, J. Chem. Phys. **49**, 4252 (1968).

[11] E. P. Wigner, On the Behavior of Cross Sections Near Thresholds, Phys. Rev. **73**, 1002 (1948).

[12] D. Hanstorp, C. Bengtsson, and D. J. Larson, Angular distributions in photodetachment from $O^-$, Phys. Rev. A **40**, 670 (1989).

[13] L. Velarde, T. Habteyes, E. R. Grumbling, K. Pichugin, and A. Sanov, Solvent resonance effect on the anisotropy of $NO^-(N_2O)_n$ cluster anion photodetachment, J. Chem. Phys. **127**, 084302 (2007).

[14] M. Van Duzor, F. Mbaiwa, J. Wei, T. Singh, R. Mabbs, A. Sanov, S. J. Cavanagh, S. T. Gibson, B. R. Lewis, and J. R. Gascooke, Vibronic coupling in the superoxide anion: the vibrational dependence of the photoelectron angular distribution, J. Chem. Phys. **133**, 174311 (2010).

[15] D. Khuseynov, C. C. Blackstone, L. M. Culberson, and A. Sanov, Photoelectron angular distributions for states of any mixed character: An experiment-friendly model for atomic, molecular, and cluster anions, J. Chem. Phys. **141**, 124312 (2014).

[16] C. C. Blackstone, A. A. Wallace, and A. Sanov, Photoelectron angular distributions in photodetachment from polarised *d* -like states: the case of $HO_2^-$, Mol. Phys. **119**, e1831636 (2021).

[17] B. Ru, C. A. Hart, R. Mabbs, S. Gozem, A. I. Krylov, and A. Sanov, Dipole effects in the photoelectron angular distributions of the sulfur monoxide anion, Phys. Chem. Chem. Phys. **24**, 23367 (2022).

[18] C. A. Hart, J. Lyle, J. Spellberg, A. I. Krylov, and R. Mabbs, Role of the Electron–Dipole Interaction in Photodetachment Angular Distributions, J. Phys. Chem. Lett. **12**, 10086 (2021).

[19] R. Mabbs, F. Mbaiwa, J. Wei, M. Van Duzor, S. T. Gibson, S. J. Cavanagh, and B. R. Lewis, Observation of vibration-dependent electron anisotropy in $O_2^-$ photodetachment, Phys. Rev. A **82**, 011401 (2010).

[20] T. Kato, On the eigenfunctions of many-particle systems in quantum mechanics, Commun. Pure Appl. Math. **10**, 151 (1957).

[21] C.-O. Almbladh and U. von Barth, Exact results for the charge and spin densities, exchange-correlation potentials, and density-functional eigenvalues, Phys. Rev. B **31**, 3231 (1985).

[22] C. G. Ning, Electron Affinities of Atoms and Structures of Atomic Negative Ions, J. Phys. Chem. Ref. Data **51**, 021502 (2022).

[23] R. L. Tang, X. L. Chen, X. X. Fu, H. Wang, and C. G. Ning, Electron affinity of the hafnium atom, Phys. Rev. A **98**, 020501 (2018).

[24] R. L. Tang, R. Si, Z. J. Fei, X. X. Fu, Y. Z. Lu, T. Brage, H. Liu, C. Chen, and C. Ning, Observation of electric-dipole transitions in the laser-cooling candidate $Th^-$ and its application for cooling antiprotons, Phys. Rev. A **103**, 042817 (2021).

[25] B. Dick, Inverting ion images without Abel inversion: maximum entropy reconstruction of velocity maps, Phys. Chem. Chem. Phys. **16**, 570 (2014).

[25] A. Szabo and N. S. Ostlund, *Modern Quantum Chemistry: Introduction to Advanced Electronic Structure Theory* (Dover Publications, Mineola, N.Y, 1996), p. 123.

[27] T. Lu and F. Chen, Multiwfn: A multifunctional wavefunction analyzer, J. Comput. Chem. **33**, 580 (2012).

[28] W. Humphrey, A. Dalke, and K. Schulten, VMD: Visual molecular dynamics, J. Mol. Graph. **14**, 33 (1996).

[29] H. W. Thompson and B. A. Green, The fundamental vibration band of nitric oxide, Spectrochim. Acta **8**, 129 (1956).

[30] P. Linstrom, *NIST Chemistry WebBook, NIST Standard Reference Database 69*, https://doi.org/10.18434/T4D303.

[31] B. Metz, H. Stoll, and M. Dolg, Small-core multiconfiguration-Dirac–Hartree–Fock-adjusted pseudopotentials for post-*d* main group elements: Application to PbH and PbO, J. Chem. Phys. **113**, 2563 (2000).

[32] K. A. Peterson, Systematically convergent basis sets with relativistic pseudopotentials. I. Correlation consistent basis sets for the post- *d* group 13–15 elements, J. Chem. Phys. **119**, 11099 (2003).

[33] K. A. Peterson, D. Figgen, E. Goll, H. Stoll, and M. Dolg, Systematically convergent basis sets with relativistic pseudopotentials. II. Small-core pseudopotentials and correlation consistent basis sets for the post- *d* group 16–18 elements, J. Chem. Phys. **119**, 11113 (2003).

[34] R. Johnson, *Computational Chemistry Comparison and Benchmark Database, NIST Standard Reference Database 101*, https://doi.org/10.18434/T47C7Z.

[35] A. R. Hoy, J. W. C. Johns, and A. R. W. McKellar, Stark Spectroscopy with the CO Laser: Dipole Moments, Hyperfine Structure, and Level Crossing Effects in the Fundamental Band of NO, Can. J. Phys. **53**, 2029 (1975).